\documentclass[aps,prb,twocolumn,superscriptaddress,showpacs]{revtex4-1}

\usepackage{graphicx}
\usepackage{calc}
\usepackage{bm}
\usepackage{color}
\usepackage{textcomp}

\begin{document}

\title{Interface superconductivity in a type-II Dirac semimetal NiTe$_2$}

\author{V.D.~Esin}
\author{O.O.~Shvetsov}
\author{A.V.~Timonina}
\author{N.N.~Kolesnikov}
\author{E.V.~Deviatov}

\affiliation{Institute of Solid State Physics of the Russian Academy of Sciences, Chernogolovka, Moscow District, 2 Academician Ossipyan str., 142432 Russia}

\date{\today}

\begin{abstract}
  We experimentally investigate charge transport through a single planar junction between a NiTe$_2$ Dirac semimetal and a normal gold lead. At millikelvin temperatures we observe non-Ohmic $dV/dI(V)$ behavior resembling Andreev reflection at a superconductor -- normal metal interface, while NiTe$_2$ bulk remains non-superconducting. The conclusion on superconductivity is also supported by suppression of the effect by temperature and magnetic field. In analogy with the known results for Cd$_3$As$_2$ Dirac semimetal,  we connect this behavior with interfacial superconductivity  due to the flat-band formation at the Au-NiTe$_2$ interface. Since the flat-band and topological surface states are closely connected, the claim on the flat-band-induced superconductivity is also supported by the Josephson current through the topological surface states on the pristine NiTe$_2$ surface. We demonstrate the pronounced Josephson diode effect, which results from the momentum shift of topological surface states of NiTe$_2$ under an in-plane magnetic field. 
\end{abstract}


\maketitle

\section{Introduction}

A  search of topological superconductivity is one of the topics that supports an interest to Dirac materials. As a most famous example, superconductivity has been observed in twisted bilayer graphene~\cite{graphene1,graphene2,graphene3}. Flat-band  formation is considered as the favorite explanation for these intriguing results~\cite{barash,kopnin2,flatTc1,flatTc2}.  For  stacking of graphene layers,  the nodal line is formed in the bulk, which is the source of the topological protection of the surface band. The two-dimensional surface flat-bands are formed from the zero energy states on the top and bottom surfaces of this artificial nodal line semimetal. The boundaries of the flat bands are the projections of the nodal loop in bulk to the top and bottom surfaces. In the presence of attractive interaction due to electron-phonon coupling, the extremely singular density of states associated with the flat band dramatically increases the superconducting transition temperature~\cite{volovik,volovik1}.

Apart from graphene, topological semimetals are also characterized by Dirac spectrum. The previous considerations on the surface flat-band can be naturally extended for nodal-line semimetals~\cite{volovik,volovik1}. However, they are also valid  for other topological semimetals~\cite{volovik3}, since the flat-band and topological surface states are closely connected: as supported by the topological-insulator-multilayer model, stacking of layers of topological insulator leads to formation of a semimetal with Fermi arc surface states~\cite{armitage}, so the flat-band is the topological surface state at zero energy~\cite{kopnin2,volovik1}. The flat band remains even if the nodal line extends, reaches the boundaries of the Brillouin zone and disappears there. In this case the nodal line semimetal transforms to the 3D topological insulator, and the surface flat bands are extended to the whole 2D Brillouin zones on the top and bottom surfaces.~\cite{volovik}.

For Cd$_3$As$_2$ Dirac semimetal,  flat bands are evidenced in angle-resolved photoemission spectroscopy (ARPES)~\cite{neu,roth}  and magneto-optic~\cite{hakl,akrap} experiments. Surface superconductivity has been experimentally observed in direct transport experiments~\cite{cdas,aronzon}. Also, point contact spectroscopy experiments~\cite{tpc1,tpc2} reveal signatures of superconductivity in a tip contact region (so-called tip induced superconductivity), while the  pressure of a tip is obviously not enough to induce bulk superconductivity~\cite{pressure} in Cd$_3$As$_2$. Due to the different experimental techniques, these results require a general explanation, so the  flat-band-stimulated superconductivity approach has an advantage of its independence of the  experimental details. 

Due to the topological origin, the effect should be also independent on the particular material. 
NiTe$_2$ is a recently discovered Dirac semimetal belonging to the family of transition metal dichalcogenides. Nontrivial topology of NiTe$_2$ single crystals has been confirmed by spin-resolved ARPES~\cite{PhysRevB.100.195134, Mukherjee2020}. 

 Though bulk NiTe$_2$ single crystals have finite resistivity down to mK temperature, the effect of the topological surface states on superconductivity has been demonstrated as Josephson diode effect~\cite{JDE}. In general, Cooper pairs can acquire a finite momentum and give rise to a diode effect in superconductors with strong spin-orbit coupling~\cite{JDE16,JDE17,JDE18}. In NiTe$_2$, the finite momentum pairing results from the momentum shift of topological surface states under an in-plane magnetic field due to the spin-momentum locking, as confirmed by ARPES measurements~\cite{JDE}. Since the flat-band and topological surface states are closely connected, it is reasonable to search also for the interface superconductivity in this type-II Dirac semimetal.

Here we experimentally investigate charge transport through a single planar junction between a NiTe$_2$ Dirac semimetal and a normal gold lead. At millikelvin temperatures we observe non-Ohmic $dV/dI(V)$ behavior resembling Andreev reflection at a superconductor -- normal metal interface, while NiTe$_2$ bulk remains non-superconducting. In analogy with the known results for Cd$_3$As$_2$ Dirac semimetal,  we connect this behavior with interfacial superconductivity  due to the flat-band formation at the Au-NiTe$_2$ interface. Since the flat-band and topological surface states are closely connected, the claim on the flat-band-induced superconductivity is also supported by the pronounced Josephson diode effect on the pristine NiTe$_2$ surface.

\section{Samples and technique}

\begin{figure}
\includegraphics[width=\columnwidth]{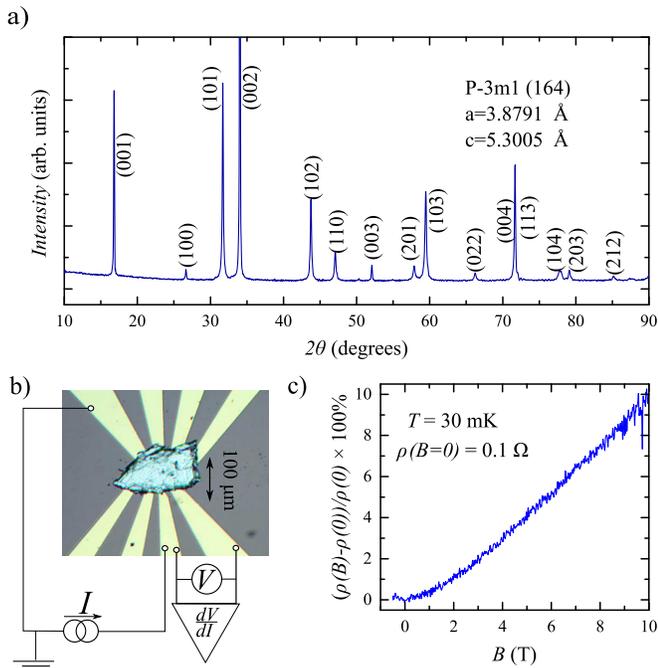}
\caption{(Color online) (a) X-ray powder diffraction pattern, which confirms single-phase NiTe$_2$ with P-3m1 (164) space group (a = b = 3.8791 \AA, c = 5.3005 \AA). (b) A top-view image of the sample with Au leads. A thick (0.5~$\mu$m) NiTe$_2$ mechanically exfoliated flake is placed on the  pre-defined Au leads pattern to form 5~$\mu$m separated NiTe$_2$-Au junctions with about 10~$\mu$m lateral size. To investigate a single NiTe$_2$-Au junction, electron transport is measured in a standard three-point technique.  (c) The known non-saturating magnetoresistance for NiTe$_2$   material~\cite{Xu2018,PhysRevB.99.155119} is reproduced for our samples in four-point longitudinal magnetoresistance measurements. The four-point resistance is finite ($\approx 0.1 \Omega$) even in zero field.  The data are obtained at $T$~= 30~mK for the normal magnetic field orientation. }
\label{fig1}
\end{figure}

NiTe$_2$ was synthesized from elements, which were taken in the form of foil (Ni) and pellets (Te). The mixture was heated in an evacuated silica ampule up to 815$^{\circ}$ C with the rate of 20 deg/h, the ampule was kept at this temperature for 48 h. The layered single crystal was grown in the same ampule by the gradient freezing technique with the cooling rate of 10 deg/h. 

The powder X-ray diffraction analysis (Cu K$\alpha$1 radiation, $\lambda$ = 1.540598 \AA) confirms single-phase NiTe$_2$ with P-3m1 (164) space group (a = b = 3.8791 \AA, c = 5.3005 \AA), see Fig.~\ref{fig1}(a).  The known structure model  is also refined with single crystal X-ray diffraction measurements (Oxford diffraction Gemini-A, Mo K$\alpha$). Nearly stoichiometric ratio Ni$_{1-x}$Te$_2$ ($x$ $<$ 0.06) is verified by the energy-dispersive X-ray spectroscopy. The obtained crystal is of 80~mm length and 5~mm thickness, with  with (0001) cleavage plane. 

Fig.~\ref{fig1}(b) shows a top-view image of a sample. Despite NiTe$_2$ can be thinned down to  two-dimensional monolayer samples, topological semimetals are essentially three-dimensional objects~\cite{armitage}. Thus, we have to select  relatively thick (above 0.5~$\mu$m) NiTe$_2$ single crystal flakes, which also ensures sample homogeneity. Thick flakes requires special contact preparation technique: the fresh mechanically exfoliated flake is transferred on the Au leads pattern, which is defined on the standard oxidized silicon substrate by lift-off technique, as depicted in Fig.~\ref{fig1}(b).  The transferred flake is shortly pressed  to the leads by another oxidized silicon substrate, the latter is removed afterward. The 100~nm thick, 10~$\mu$m wide Au leads are separated by 5~$\mu$m intervals under the flake.  This procedure provides transparent Au-NiTe$_2$ junctions (about 2~Ohm resistance for the 10~$\mu$m lateral size), stable in different cooling cycles, which has been verified  before for a wide range of materials~\cite{cdas,inwte1,inwte2,incosns,black,timnal,infgt}. As an additional advantage, the obtained Au-NiTe$_2$ junctions are protected from any contamination by SiO$_2$ substrate.

The quality of our NiTe$_2$ material can be tested in standard four-point magnetoresistance measurements, see Fig.~\ref{fig1} (c). NiTe$_2$ flakes demonstrate  non-saturating longitudinal magnetoresistance for the normal magnetic field orientation, which well reproduces the previously  reported one for this material~\cite{Xu2018,PhysRevB.99.155119}. The four-point resistance is finite (0.1~$\Omega$) between 5~$\mu$m spaced Au leads, so there is no superconductivity for bulk NiTe$_2$ single crystal flakes at ambient pressure.   The measurements are performed within the 30~mK -- 1.2~K temperature range in a dilution refrigerator equipped with a superconducting solenoid.

\section{Experimental results}

\subsection{Single Au-NiTe$_2$ junctions}

\begin{figure}
\includegraphics[width=\columnwidth]{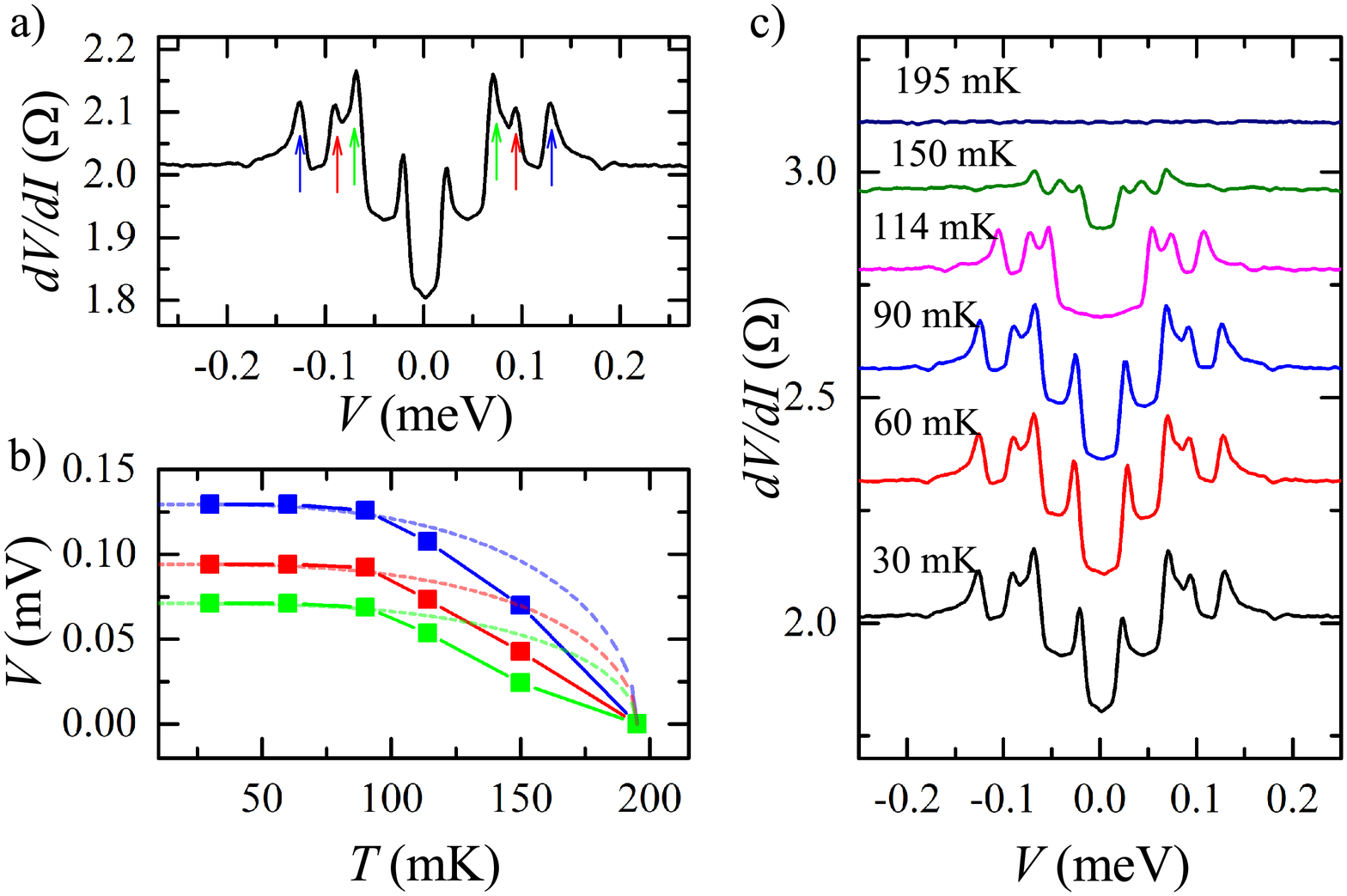}
\caption{(Color online) (a) Non-Ohmic behavior of $dV/dI(V)$ differential resistance for a single Au-NiTe$_2$ junction at 30~mK temperature in zero magnetic field. The $dV/dI(V)$ curve shows prominent  $dV/dI$ drop around the zero bias, which is accompanied by several $dV/dI$ peaks at higher biases (denoted by arrows for the positive bias polarity).  (b) Temperature dependence of the peaks' positions. The colors correspond to the arrow colors in the (a) panel. The standard BCS fit~\cite{tinkham} is shown by the dashed lines. (c) Temperature dependence of $dV/dI(V)$ curves in zero magnetic field. The central $dV/dI$ drop is diminishing with the temperature, while  $dV/dI$ peaks' positions also move to the zero bias. The curves are shifted for clarity.}
\label{fig2}
\end{figure}

We study electron transport across a single Au-NiTe$_2$ junction in a standard three-point technique: one Au contact is grounded and two other contacts are used as current and voltage probes, as schematically presented in Fig.~\ref{fig1}(b). To obtain $dV/dI(V)$ characteristics in Fig.~\ref{fig2} (a), dc current is additionally modulated by a low (100~nA) ac component. We measure both dc ($V$) and ac ($\sim dV/dI$) voltage components  with a dc voltmeter and a lock-in amplifier, respectively. The signal is confirmed to be independent of the modulation frequency within 100 Hz -- 10kHz range, which is defined by the applied filters.

Fig.~\ref{fig2} (a)  shows non-Ohmic behavior of $dV/dI(V)$ differential resistance for a single Au-NiTe$_2$ junction. We observe the prominent  ( about 10\%)  $dV/dI$ drop around the zero bias, which is accompanied by several $dV/dI$ peaks at higher biases. For a fixed grounded Au contact,  $dV/dI(V)$ curve is verified to be independent of mutual positions of current/voltage probes, so it mostly  reflects the resistance of Au-NiTe$_2$ interface (about 2~$\Omega$) without noticeable admixture of the sample’s bulk ($\approx 0.1 \Omega$ resistance in Fig.~\ref{fig1} (c)).

Fig.~\ref{fig2} (b,c) shows $dV/dI(V)$ temperature dependence. Non-Ohmic behavior can only be seen below the $T_c \approx$ 190~mK critical temperature. The central $dV/dI$ drop is diminishing with the temperature, while  $dV/dI$ peaks' positions also move to the zero bias. The latter dependence is directly depicted in  Fig.~\ref{fig2} (b) for three successive peaks from Fig.~\ref{fig2} (a).

\begin{figure}
\includegraphics[width=\columnwidth]{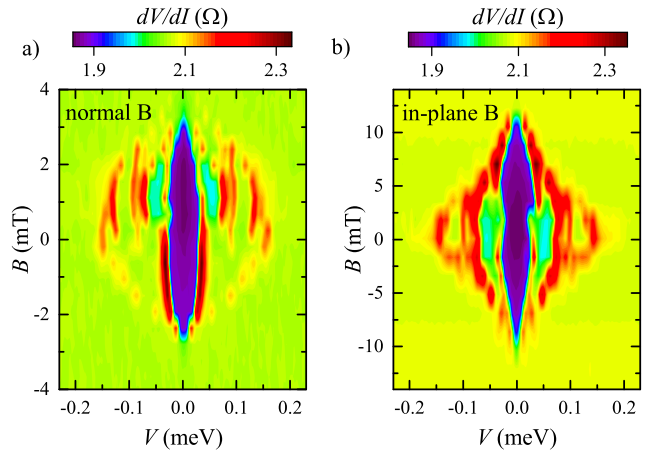}
\caption{(Color online) Suppression of $dV/dI(V)$ non-Ohmic behavior by magnetic field at 30~mK temperature for normal (a)  and in-plane (b) field orientations. The behavior is qualitatively similar, while the critical fields $B_c$ differ significantly for these two orientations. There is also some asymmetry of the colormap for normal field orientation (a), which is verified to be independent on the magnetic field sweep direction.  }
\label{fig3}
\end{figure}

\begin{figure}
\includegraphics[width=\columnwidth]{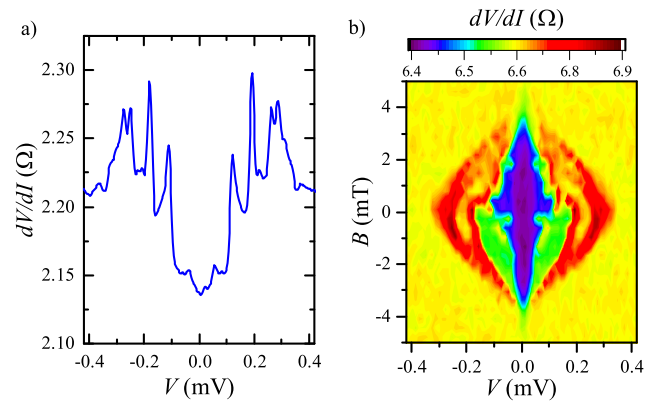}
\caption{(Color online) (a) Non-Ohmic behavior of $dV/dI(V)$  for a different sample to show typical sample-to-sample variation. The behavior is qualitatively the same as for the first junction in Fig.~\ref{fig2}. (b) Suppression of $dV/dI(V)$ curves in normal magnetic field.  Some $dV/dI(V,B)$ asymmetry can also be seen for this sample. The data are obtained at 30~mK temperature. }
\label{fig4}
\end{figure}

Non-Ohmic behavior can also be suppressed by magnetic field, see Fig.~\ref{fig3} (a) and (b) for normal and in-plane field orientations, respectively. Despite of the qualitative similarity, the critical fields differ significantly for these two orientations:  $B_c$ can be estimated as 3~mT in Fig.~\ref{fig3} (a) and as 10~mT in Fig.~\ref{fig3} (b). We also observe some asymmetry of the colormap in Fig.~\ref{fig3} (a) for normal magnetic field, which can not be seen for in-plane orientation in Fig.~\ref{fig3} (b). The normal-field asymmetry is verified to be independent on the magnetic field sweep direction.  The asymmetry can not be connected with magnetic ordering in the bulk of NiTe$_2$, since mangetometry revealed a purely paramagnetic susceptibility for our NiTe$_2$ crystals in accordance with the previously reported data~\cite{Xu2018}.

We observe similar non-Ohmic $dV/dI$ behavior for several Au-NiTe$_2$ junctions, the critical temperature of the non-linearity suppression is varied within the 150-200~mK range from sample to sample.  Fig.~\ref{fig4} represents one of the examples. $dV/dI(V)$ non-linearity is qualitatively the same in Fig.~\ref{fig4} (a)  as for the first junction in Fig.~\ref{fig2} (a). Some $dV/dI(V,B)$ asymmetry can also be seen for this sample in normal magnetic fields, as depicted in Fig.~\ref{fig4} (b).

The drop in  $dV/dI(V)$ can not be attributed to the usual scattering at the Au-NiTe$_2$ interface, since the scattering should be described as the effective interface  potential and, therefore, it always results in a wide $dV/dI$ peak at zero bias~\cite{Rajput2013}. On the other hand, $dV/dI(V)$ continuous increase to both sides of zero bias is known for  electron-phonon or electron-magnon scattering~\cite{Myers1999,Duif1989}, but the temperature and bias voltage ranges  apparently refer to a much smaller energy scales in our experiment.

The observed behavior strongly resembles the known one for typical Andreev reflection at the NS interface between a normal metal and a superconductor~\cite{tinkham,AndreevReflection}. In this case, multiple $dV/dI(V)$ peaks in Fig.~\ref{fig2} (a) should be attributed to geometrical resonances in NSN junctions~\cite{osbite,tomasch1,tomasch2}. In the case of a thin superconducting layer (S), these resonances are known as crossed Andreev reflection, where an incident electron and a reflected hole appear to both sided of the S layer~\cite{CAR}. The resonance positions are determined by the superconducting gap, so they follow the gap temperature  dependence in Fig.~\ref{fig2} (b). 

The standard BCS temperature dependence~\cite{tinkham} is shown by the dashed lines in Fig.~\ref{fig2} (b) for several peak positions. The width of the central  $dV/dI$ drop well corresponds to the temperatures of the $dV/dI(V)$ curves smearing in Figs.~\ref{fig2} and~\ref{fig4}.  The claim on the superconductivity is also confirmed by magnetic field suppression of the effect. For the planar experimental geometry, it is natural to have the critical field anisotropy for normal and in-plane field orientations in Fig.~\ref{fig3} (a) and (b), respectively.

\subsection{Double Nb-NiTe$_2$-Nb junctions}

The claim on the interface superconductivity can be further supported by the supercurrent investigations between two Nb-NiTe$_2$ interfaces. We study electron transport between two neighbor $1 \mu$m Nb leads in a standard two-point technique. All the wire resistances are excluded, which is necessary for low-impedance samples. 

\begin{figure}
\includegraphics[width=\columnwidth]{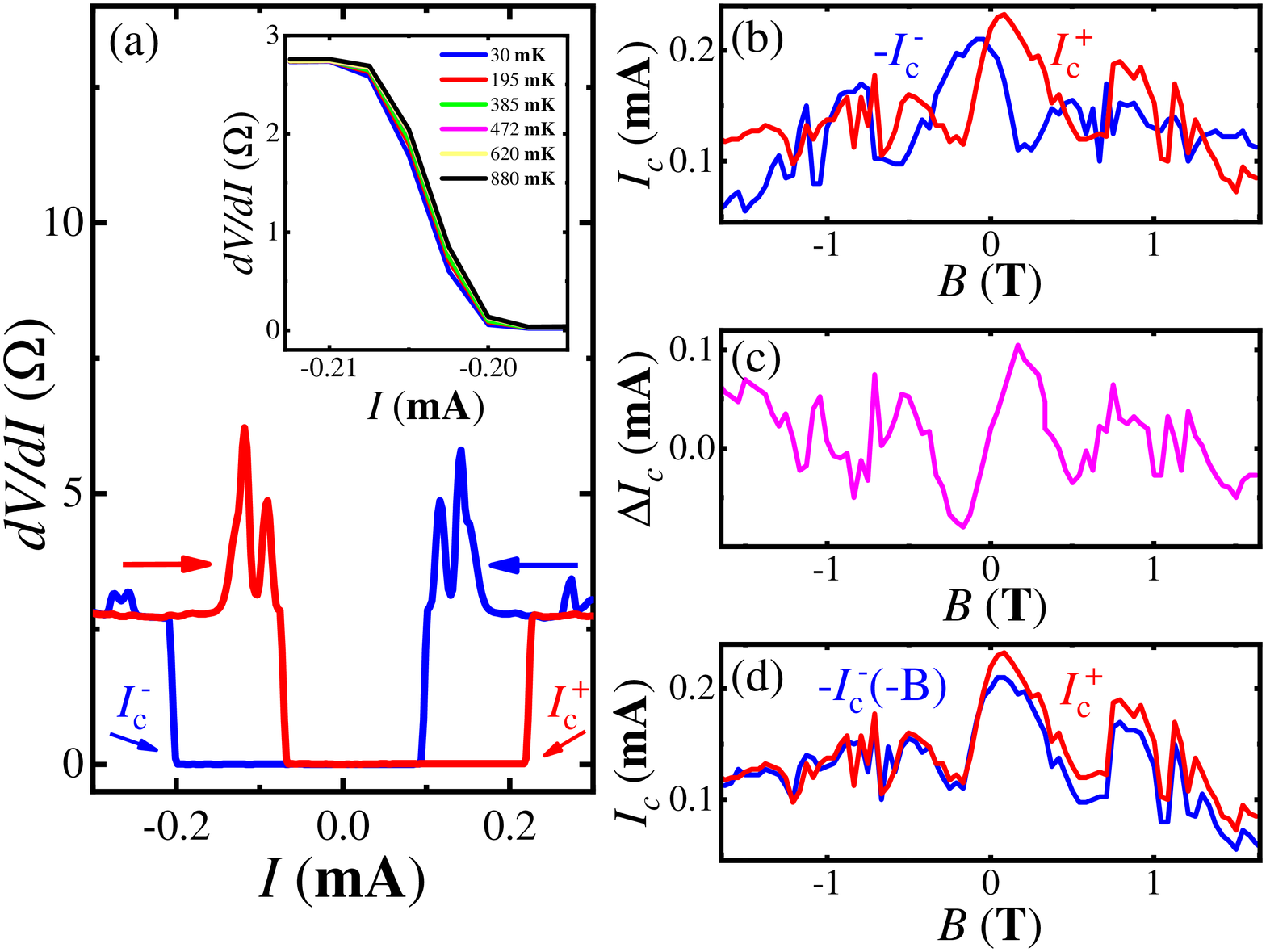}
\caption{(Color online) Josephson diode effect for a double Nb-NiTe$_2$-Nb junction. (a)  Examples of $dV/dI(V)$ curves for two different current sweep directions for 26~mT in-plane magnetic field. Apart from standard hysteresis for superconducting-resisitve and resisitive-superconducting transitions,  non-reciprocal response is seen even for the same superconducting-resistive part of the curves,  as depicted by $I_c^+=0.22$~mA and $I_c^-=-0.20$~mA values. Inset shows temperature stability of $dV/dI(V)$ curves in the 30~mK -- 1.2~K range. (b)  Critical currents  $I_c^+(B)$ (red) and $-I_c^-(B)$ (blue) for two different current sweep directions, respectively, in dependence of the in-plane magnetic field. (c) The difference  $\Delta I_c = I_c^+ -I_c^-$, which is nearly antisymmetric in magnetic field, with multiple sign changes~\cite{JDE}. (d) Another demonstration of the field antisymmetry, as coincidence of $I_c^+(B)$ and the reversed $-I_c^-(-B)$ curves, which also confirms high accuracy of $I_c$ determination.}
\label{fig5}
\end{figure}

Fig.~\ref{fig5} (a)  clearly demonstrates Josephson $dV/dI(I)$ curves for two different current sweep directions. As expected, the zero-resistance state appears below some critical current $I_c$. The absolute value of  $I_c$ is strongly different for resistive-superconducting and superconducting-resisitve transitions, so $dV/dI(I)$ curves show standard hysteresis with current sweep direction. Also, the transition region  is nearly independent of temperature in the 30~mK -- 1.2~K range, because of high critical temperature on niobium, see the inset to Fig.~\ref{fig5} (a).

However,  the absolute $I_c$ values are also different for two current sweep directions even for the same superconducting-resistive part of the curves, as depicted by $I_c^+=0.22$~mA and $I_c^-=-0.20$~mA values in  Fig.~\ref{fig5} (a) for +26~mT  in-plane magnetic field. This non-reciprocal response is known as the Josephson diode effect, the simplest manifestations of which is the direction dependence of the critical current.

To obtain $I_c$ with high accuracy at fixed $B$, we sweep the current ten times from the zero value (i.e. from the superconducting $dV/dI$ = 0 state) to some value well above the $I_c$ (the resistive $dV/dI > 0$ state) and then determine $I_c$ as an average value of $dV/dI$ breakdown positions. 

The result is presented in Fig.~\ref{fig5} (b) as $I_c^+$ (red) and $-I_c^-$ (blue) for two different current sweep directions, respectively.  The difference $\Delta I_c = I_c^+ -I_c^-$ is governed by magnetic field, demonstrating odd-type field dependence in Fig.~\ref{fig5} (c), as expected for the Josephson diode effect. This can also be demonstrated by direct comparison of $I_c^+(B)$ and $-I_c^-(-B)$ in Fig.~\ref{fig5} (d). The curves coincide well, they show somewhat distorted Fraunhofer pattern~\cite{aperiodicsfs1,aperiodicsfs2}. Josephson diode effect is due to the spin-momentum locking in topological Dirac semimetal NiTe$_2$, which connects the superconductivity and the topological surface states.  

\section{Discussion}

As a result, $dV/dI(V)$ curves for an individual Au-NiTe$_2$ interface strongly resemble the effect  of the interface superconductivity in NiTe$_2$ Dirac semimetal. Similar result was previously observed~\cite{cdas,aronzon} for another Dirac semimetal Cd$_3$As$_2$, which  was attributed to the flat band formation at the Au-Cd$_3$As$_2$ interface~\cite{cdas}. Thus, the interface superconductivity should  reflect the fundamental  physics of topological Dirac semimetals, irrespective of the specifics of the  particular material. 

As an opposite example of material-dependent effects, bulk superconductivity  is known for pressurized Te-deficient NiTe$_2$~\cite{Feng2021} as well as for doped~\cite{wang1}  Cd$_3$As$_2$ single crystal samples.  These effects can be ruled out in our experiment, since i) bulk superconductivity is not observed for our NiTe$_2$ crystals according to four-point resistance data in Fig.~\ref{fig1}(c); ii) X-ray spectroscopy reveals almost stoichiometric Ni$_{1-x}$Te$_2$  crystal with a slight Ni deficiency ($x$~$<$ 0.06); iii) there is no external pressure in our experiment.

On the other hand, interface superconductivity  can appear due to the flat-band formation~\cite{barash,kopnin2,flatTc1,flatTc2}, which is the topological phenomenon~\cite{volovik,volovik1,volovik3}. In Dirac semimetals, strain generically acts as an effective gauge field on Dirac fermions and creates pseudo-Landau orbitals without breaking time-reversal symmetry~\cite{strain}. The zero-energy Landau orbitals form a flat band in the vicinity of the Dirac point, so the high density of states of this flat band may produce the interface superconductivity.

Strain can occur at the interface between Au and NiTe$_2$ due to the lattice mismatch. The strain-induced flat-band formation is predicted in pristine NiTe$_2$ at ambient pressure~\cite{PhysRevB.103.125134}, so the statement on the Au-NiTe$_2$ interface superconductivity is quite  reasonable.  It is important, that we observe finite four-point resistance between different contacts in Fig.~\ref{fig1}(c), which well correspond to the fact that strain-induced flat-band formation is only occurs   at  Au-NiTe$_2$ interface. 

In topological materials, the flat-band and topological surface states are closely connected: they appear due to the bulk-boundary correspondence, so the flat-band is the topological surface state at zero energy~\cite{kopnin2,volovik1}.  The topological surface states are essentially spin-polarized~\cite{PhysRevB.100.195134} due to the spin-momentum locking. It may result to the $dV/dI(V,B)$ asymmetry, observed for Au-NiTe$_2$ junctions subjected to the normal magnetic fields (see Fig.~\ref{fig3}(a) and~\ref{fig4}(b)). Which is more important, spin-momentum locking is responsible for the Josephson diode effect. 

The Josephson diode effect appears as the direction-dependent Josephson current, where the direction of the Cooper pair momentum determines the polarity of the effect. In Fig.~\ref{fig5}, the finite Cooper pair momentum appears as  antisymmetric  $\Delta I_c (B)$ dependence.  The finite momentum pairing results from the momentum shift of topological surface states of NiTe$_2$ under an in-plane magnetic field, so Josephson diode effect originates from spin-helical topological surface states, in an otherwise centrosymmetric system. Thus, the  Josephson diode effect connects the superconductivity and the topological surface states in NiTe$_2$. 

As a result, topological surface states carry the Josephson current on the pristine NiTe$_2$ surface in between the two neighbor metallic contacts, while these states are responsible for the flat-band formation at the  Au-NiTe$_2$ interface, and, therefore,  the interface superconductivity.

\section{Conclusion}

As a conclusion, we  experimentally investigate charge transport through a single planar junction between a NiTe$_2$ Dirac semimetal and a normal gold lead. At millikelvin temperatures we observe non-Ohmic $dV/dI(V)$ behavior resembling Andreev reflection at a superconductor -- normal metal interface, while NiTe$_2$ bulk remains non-superconducting. We connect this behavior with interfacial superconductivity  due to the flat-band formation at the Au-NiTe$_2$ interface. Also, 
 we demonstrate the pronounced Josephson diode effect on the pristine NiTe$_2$ surface, which results from the momentum shift of topological surface states under an in-plane magnetic field. This observation further supports the claim on the flat-band-induced superconductivity, since the flat-band and topological surface states are closely connected.

\acknowledgments
We wish to thank G.E.~Volovik and Yu.S.~Barash  for fruitful discussions, and S.S.~Khasanov for X-ray sample characterization. We gratefully acknowledge financial support  by the  Russian Science Foundation, project RSF-22-22-00229, https://rscf.ru/project/22-22-00229/.

\end{document}